\documentclass[11pt,cits]{JHEP3}
\usepackage{epsfig}
\usepackage{latexsym}
\usepackage{graphicx}
\usepackage{amsmath,amssymb}
\usepackage{epstopdf}  
\def\unit{\relax{\rm 1\kern-.26em I}}

\def\bea{\begin{eqnarray}}
\def\eea{\end{eqnarray}}
\def\be{\begin{equation}}
\def\ee{\end{equation}}
\def\nn{\nonumber}

\newcommand{\gsim}{\lower.7ex\hbox{$\;\stackrel{\textstyle>}{\sim}\;$}}
\newcommand{\lsim}{\lower.7ex\hbox{$\;\stackrel{\textstyle<}{\sim}\;$}}

\title{An Effective Description of the Landscape -- II}
\author{ \\
\\
\email{}}
\author{Diego Gallego and Marco Serone \\
International School for Advanced Studies (SISSA/ISAS)
and INFN, Trieste, Italy\\
E-Mail: \email{gallego@sissa.it,serone@sissa.it}}

\abstract{We continue our analysis of establishing the reliability of ``simple" effective theories where massive fields are  ``frozen''  rather than integrated out, in  a wide class of four dimensional theories with global or local  ${\cal N}=1$ supersymmetry.  We extend our previous work by adding gauge fields and ${\cal O}(1)$ Yukawa-like terms for the charged fields in the superpotential.
For generic K\"ahler potentials, a meaningful freezing is allowed for chiral multiplets only, whereas
in general heavy vector fields have to properly be integrated out.
Heavy chiral fields can be frozen if they approximately sit to supersymmetric solutions along their directions and, in supergravity, if the superpotential at the minimum is small, so that a mass hierarchy between heavy and light fields is ensured. When the above conditions are met, we show that  the simple effective theory is generally  a reliable truncation of the full one.}

\preprint{SISSA-19/2009/EP}

\keywords{Supergravity Models, Supersymmetry Breaking, dS vacua in string theory}

\begin{document}

\section{Introduction}

Effective field theories arising from string compactifications are notoriously difficult to study, due to the proliferation of fields that one generally gets.
The first guiding principle to understand the impact of a given state on the low-energy dynamics is its mass. Whenever the latter is sufficiently heavy, we expect
the effect of the related state to be small enough that we can safely neglect it. 
A very common and drastic simplification is to assume that 
string, Kaluza-Klein and winding resonances are all so heavy that they can be safely neglected.
In so doing, we are left with a given supergravity (SUGRA) field theory, assumed from now on to be four-dimensional with ${\cal N}=1$ supersymmetry (SUSY). Despite this huge simplification, an explicit study of such SUGRA theories is still a formidable task, containing typically hundreds of fields.
Even just finding the vacua is often a hard task.
The  recent progress in string compactifications with fluxes has shown that many of these fields can get
large masses,\footnote{By consistency of the construction, such masses should however be smaller than 
the lightest  string, Kaluza-Klein and winding states.} so that one can integrate them out and simplify the low energy theory, in the spirit of effective theories. However, the multiplicity of these fields is such that  even at the classical level  integrating them out is practically very hard. As a matter of fact, the typical attitude taken in the literature consists in just neglecting these fields. 

In a previous companion paper \cite{LaI}, we have started to study, in a general ${\cal N}=1$ supersymmetric setting,  under what conditions massive fields can be neglected rather than integrated out, in theories with chiral multiplets 
only.\footnote{See \cite{deAlwis:2005tg,GomezReino:2006dk,GomezReino:2007qi,Achucarro:2007qa} for previous related studies of this sort and \cite{Choi:2004sx,deAlwis:2005tf,Abe:2006xi,Choi:2008hn} for studies in the context of the Kachru-Kallosh-Linde-Trivedi (KKLT)  \cite{Kachru:2003aw} and related scenarios.}  Since the notion itself of heavy and light fields manifestly depends on the vacuum, the requirement to be able to uniquely define heavy fields
in a wide regime in field space for the remaining light states, forced us to study SUSY theories with a superpotential of the form
\be
W(H,L) = W_0(H) + \epsilon \, W_1(H,L)\,, 
\label{Wint}
\ee
with $\epsilon \simeq m_L/m_H \ll 1$ being the ratio of the typical light over heavy mass scale, $H$ and $L$ denoting schematically heavy and light  chiral superfields. 
Our ansatz for $W$  is also motivated by the fact that several recent string compactifications admit a  superpotential of the form (\ref{Wint}). For instance, in the well-known class of Type IIB flux compactification models \cite{Giddings:2001yu}, $W_0$ can be identified with the Gukov-Vafa-Witten superpotential \cite{Gukov:1999ya}, the heavy fields $H$ are identified with the dilaton and the complex structure moduli of the underlying Calabi-Yau manifold, and the $L$ with the K\"ahler structure moduli. The parameter $\epsilon$ is a schematic  way to recall that all terms in $W_1$ are non-perturbatively generated and that
in a wide region in moduli space are all suppressed.

As discussed in some detail in
\cite{LaI}, assuming that all the eigenvalues of the K\"ahler metric are parametrically larger than $\epsilon$, for any K\"ahler potential mixing term between $H$ and $L$, it is always possible to define a canonically normalized field basis where the physical heavy field fluctuations are entirely given by linear combinations of the $H$.
Hence the superpotential form (\ref{Wint}) defines which are the heavy fields in the theory.
It has shown in \cite{LaI} that in theories with a superpotential of the form (\ref{Wint}), heavy fields can be neglected, but provided they are ``frozen"  to approximate SUSY values $H_0$, defined as the solutions of $\partial_H W_0 = 0$ and that,
of course, all of them effectively develop large physical mass terms from the term $W_0$ in eq.(\ref{Wint}). 
In SUGRA theories the further condition $\langle W_0 \rangle \sim {\cal O}(\epsilon)$ is needed,\footnote{The condition  $\langle W_0 \rangle \sim {\cal O}(\epsilon)$ is not invariant under K\"ahler transformations.
More precisely, the requirement is the existence of a K\"ahler gauge where $\langle W_0 \rangle \sim {\cal O}(\epsilon)$, $K\sim {\cal O}(1)$ and the K\"ahler metric has eigenvalues parametrically larger than $\epsilon$.}
ensuring that curvature terms are small enough so that the hierarchy between heavy and light fields is retained, and the mass splitting induced by SUSY breaking negligible in the heavy field directions.\footnote{ 
If two sectors in a theory are sufficiently screened,  one can effectively study the dynamics of one of them, neglecting (freezing) the other, even in absence of a hierarchy in masses between the two sectors. 
The conditions when this can happen in SUGRA theories have been given in \cite{Binetruy:2004hh} (see also \cite{Achucarro:2007qa}) and have been
also discussed in \cite{LaI} in the context of some string compactifications \cite{ Balasubramanian:2005zx} where they approximately apply.
In that case $W_0$ can be arbitrary. We do not consider in this paper this alternative possibility.}

Aim of the present paper is to extend the study of \cite{LaI} by adding gauge fields. 
In order to further extend the class of theories falling in our analysis, we also 
generalize the form of the superpotential (\ref{Wint}) by including ${\cal O}(1)$ field-dependent Yukawa couplings for the charged fields. Models of this sort necessarily imply various scales, possibly associated to 
different symmetry breaking mechanisms, SUSY breaking, gauge symmetry breaking, etc.
In order to keep our analysis as simple as possible, and yet capture the essential features, we will 
assume the presence of just two kinds of light charged fields, characterized by having Vacuum Expectation Values (VEV) parametrically larger than $\epsilon$ and of ${\cal O}(\epsilon)$. We denote them respectively by $Z$ and $C$.  The remaining light fields are denoted by $M$. The latter can also transform under gauge transformations, but only non-linearly.
They would correspond, in explicit string constructions, to light moduli fields.\footnote{When the gauge symmetry is broken, a combination of the fields $L$ actually get a heavy mass. Nevertheless, in order to distinguish them from the fields $H$ appearing in eq.(\ref{WYukCInt}), with an abuse of language, we will keep calling them light.}  Like the light fields, the heavy fields can also be standard charged  fields or moduli-like fields, but we will mostly focus on the case in which they are all moduli-like fields.
The schematic form of the superpotential in this more complicated set-up is taken as follows:
\be
W = W_0(H) + Y_N(H,M,Z) C^N  + \epsilon \Big[ W_1(H,M,Z) + \mu_M(H,M,Z)  C^M\Big] \,,
\label{WYukCInt}
\ee
where $W_0$, $Y_N$, $W_1$ and $\mu_M$ ($N\geq 3, M\geq 2$),
are arbitrary holomorphic functions, constrained only by gauge invariance. 
No assumption on the dependence of the superpotential on the charged fields $Z$ is required, so that our results can apply also in presence of non-perturbatively generated superpotential terms, 
non-polynomial in the $Z$'s. The K\"ahler potential is arbitrary, with the only assumptions that admits a Taylor expansion in the charged fields $C$ and that, as already mentioned,  all the eigenvalues of the associated K\"ahler metric are parametrically larger than  $\epsilon$. Similarly, the gauge kinetic functions $f$ are taken to be arbitrary, but regular, moduli dependent holomorphic functions.

First of all, we notice that, contrary to the fields $H$, massive gauge fields do not generally admit a freezing. More precisely, while superpotentials of the form (\ref{WYukCInt}) include a very wide class of known superpotentials arising from string compactifications, the class of K\"ahler potentials which would allow a freezing of the vector fields is quite limited and not very interesting, unless the vector field is heavy and decoupled, in which case one can trivially set it to zero. Hence we will not insist in freezing massive vector fields, but rather we will only show how the freezing of the heavy chiral fields is (not) affected in presence of heavy vector fields, the latter being always properly integrated out.

In general,  the scalar condition $F_{H,0}=\partial_H W_0=0$ does not  fix all the VEV of the heavy fields $H$, since gauge invariance constrains the form of $W_0$.  Only gauge invariant combinations of the heavy moduli, being well-defined, can reliably be frozen. The orthogonal combinations will not appear in $W_0$, but possibly in other terms of the superpotential (\ref{WYukCInt}), in combination with light fields and/or in $D$-terms. Hence, these remaining gauge invariant combinations will typically be relevant in the low-energy dynamics
and should  properly be included among the light fields.
More precisely, one can neglect non-neutral heavy moduli, assuming of having chosen a gauge-fixing
where they are gauged away, but then one has to carefully take into account the dynamics
of the associated massive vector-super field. In the context of SUSY breaking, the possible $D$-term SUSY breaking contributions hidden in the massive vector super field are generally non-negligible.
We emphasize this point because, although already made present in the literature in
various contexts \cite{ArkaniHamed:1998nu,Binetruy:2004hh}, it seems to have been overlooked in some string constructions, where gravity and moduli are neglected altogether. The impossibility of naively neglecting non-neutral moduli has nothing to do with gravity but is purely dictated by gauge invariance, so it remains also in the global limit with gravity decoupled.

Along the lines of \cite{LaI}, we compare the ``full'' effective theory obtained by classically integrating out the heavy fields to the ``simple" one obtained by just ``freezing'' them. We use a manifestly SUSY approach
in which one solves the super-field equation $\partial_H W =0$ and then plug back the result in $W$, $K$ and $f$ to get the effective full quantities $W_{full}$, $K_{full}$ and $f_{full}$. We find that in the charged field range $|C|/m_H \lesssim {\cal O}(\epsilon)$, for any values of the $M$ and $Z$ fields where the superpotential $W$ has the form (\ref{WYukCInt}), the simple theory is a reliable effective field theory. More precisely $W_{sim}$, $K_{sim}$ and $f_{sim}$ differ by the corresponding full quantities only by operators which are sub-leading in an effective field theory sense and correct the coefficients of the already existing couplings by a small amount.  The leading $C$-dependent  part of the scalar potential is identical in both theories.
On the other hand, cubic terms of the schematic form $C^3$ in $K$ and $C^6$ terms in $W$ (and higher) 
are not reliable in the simple theory, so care has to be taken in working with it anytime higher order operators are considered.
We have also checked, by working in a component approach, that higher terms in the light auxiliary fields, not detected by just solving $\partial_H W=0$, are always negligible in the scalar potential, namely they can only give rise to corrections of the same order of magnitude or smaller than the ones given by the manifestly SUSY integration of the heavy fields.

All the above results apply to global and local SUSY theories, provided that in the latter theories $\langle W_0 \rangle \sim {\cal O}(\epsilon)$, as mentioned above. This is in agreement with the recent results of \cite{Brizi:2009nn}, where it has been pointed out that the meaningful super-field equation that should be used to integrate out heavy fields is $\partial_H W =0$ in both the global and local SUSY case. It has also been emphasized in \cite{Brizi:2009nn} that the condition $W\ll 1$ is always necessary in order to ensure that the effective theory is still described by a standard SUGRA theory up to two derivative level. 
We emphasize here that the (generally) stronger condition $W\sim {\cal O}(\epsilon)$ allows a freezing of the heavy fields, under the conditions discussed above.

The structure of the paper is as follows. In section 2, after a brief review of the results of \cite{LaI}, we consider global and local SUSY theories with $W$ as in eq.(\ref{WYukCInt}),  in the limit of vanishing gauge couplings. In section 3 we include vector multiplets. We first consider theories with unbroken gauge group and then analyze theories with broken gauge group, where  heavy vector fields have to be integrated out.
We draw our conclusions in section 4. We report in Appendix A a more detailed study in components of the comparison of the equations of motion (e.o.m.) in the full and simple theory. For illustrative purposes, we report in Appendix B a numerical study of a  ``string-inspired'' SUGRA model with an $U(1)$ gauge symmetry where our considerations can be concretely applied. In order to simplify the notation, we set the the cut-off of the microscopic theory $\Lambda$ to be the reduced Planck mass  $M_p$ and use units in which $\Lambda = M_p=1$.  In addition, in order to not introduce some additional hierarchy of scales, we
assume that $m_H/M_p\gg \epsilon$ and that $m_V\sim m_H$, $m_V$ being the scale of the heavy vectors.

\section{Chiral multiplets only}

Before turning our attention to full models with gauge interactions and vector multiplets, it is useful to work in absence of the latter, in the limit of vanishing gauge couplings.
In the next subsection we first rederive the results of \cite{LaI}, valid in absence of ${\cal O}(1)$ Yukawa couplings and obtained working in field components, using a manifestly SUSY approach.

\subsection{No ${\cal O}(1)$ Yukawa couplings}\label{Noyukawa}

The study performed in \cite{LaI}, restricted to superpotentials of the form $W=W_0(H)+\epsilon W_1(H,L)$, with $\epsilon \ll 1$, established that up to ${\cal O}(\epsilon^2)$ the (full) effective scalar potential one gets by integrating out heavy moduli in a given vacuum is identical to the one obtained by simply  freezing the heavy fields at their leading (in $\epsilon$) VEV defined as $\partial_H W_0 = 0$. The results are valid in both global and local SUSY, where in the latter case the additional constraint $\langle W_0 \rangle \sim {\cal O}(\epsilon)$ is required. In this way, it has been shown in \cite{LaI} that SUSY is 
broken at ${\cal O}(\epsilon)$ by the light fields, with a suppressed back-reaction of SUSY breaking on the heavy fields, $F^H\sim {\cal O}(\epsilon^2)$. In what follows, in order to use the same notation and conventions adopted in \cite{LaI}, 
we will take $m_H\lesssim 1$  so that $m_H$ does not explicitly appear in the formulae. We will relax this
condition in the next subsection. 

Let us start by considering the global SUSY case. 
When SUSY is broken at a scale of ${\cal O}(\epsilon)$, parametrically smaller than the mass scale of the heavy fields, a manifestly SUSY approach to integrate out heavy fields is allowed. As well known, neglecting covariant and ordinary derivatives, this approach amounts in solving at the super-field level the equations 
\be
\partial_H W = 0
\label{dHW0}
\ee
and then plugging them back in $W$ and $K$ (see e.g. \cite{Affleck:1984xz}). 
In this way, one automatically integrates out the heavy fields and gets the full effective K\"ahler and superpotential $K_{full}$ and $W_{full}$. 
Strictly speaking, the effective scalar potential one obtains from $W_{full}$ and $K_{full}$
differs in general from that obtained by integrating out the heavy fields in components,  since in the latter one is exactly resumming (at zero momentum) all the auxiliary components of the light fields. In the manifestly SUSY approach with covariant derivatives neglected, instead, by definition one considers all the auxiliary components up to quadratic order only. In other words, the effective $K$ associated to the scalar potential computed in components, as in \cite{LaI}, 
is in general an infinite sum of terms involving covariant SUSY derivatives.  
In the situation at hand, however, the light fields of the $F$-terms are of ${\cal O}(\epsilon)$ and 
thus all cubic and higher $F$ terms of the light fields are negligible for our purposes.
We solve the equations $\partial_H W=0$ perturbatively in $\epsilon$:
\be
H^i = H_0^i  +  \epsilon H_1^i(L)+ {\cal O}(\epsilon^2) 
\label{deltaHexp0}
\ee
where $H_0^i$ are defined by $\partial_i W_0(H_0) = 0$ (notation as in \cite{LaI}). The effective K\"ahler and superpotential read
 \bea
 W_{full} & = & W_{sim} +  \epsilon^2 \Big(\frac 12\partial_i \partial_j W_0 H_1^i H_1^j  +\partial_i W_1H_1^i\Big) + {\cal O}(\epsilon^3) \,,  \label{WeffSUSY0} \\
 K_{full} & = &  K_{sim} +\epsilon \Big( \partial_i K_{sim}  H_1^{i}  + \partial_{\bar i} K_{sim}  
 \bar H_1^{\bar i} \Big) + {\cal O}(\epsilon^2)\,,
 \label{KeffSUSY0}
 \eea
where $W_{sim} = W(H_0)$ and $K_{sim} = K(H_0,\bar H_0)$.
The leading shift $H_1^i$ equals
 \be
H_1^i  =   -  W_0^{ij}\partial_j W_1,
\label{H1SUSY}
 \ee
with $W_0^{ij}\sim {\cal O}(1)$, the inverse of  $W_{0,ij}\equiv \partial_i \partial_j W_0$, and gives 
 \be
 W_{full}  =  W_{sim}  - \frac 12 \epsilon^2  \partial_i W_1 W_0^{ij} \partial_j W_1+ {\cal O}(\epsilon^3) \,.  \label{WeffSUSY01}
 \ee
Independently of the K\"ahler potential and its corrections, the $\epsilon^2$ terms in $W_{full}$ can only contribute to the effective scalar potential at ${\cal O}(\epsilon^3)$ and hence 
\be
V_{full} = V_{sim} + {\cal O}(\epsilon^3)\,.
\label{Vagree}
\ee
It has been recently shown that in SUGRA eq.(\ref{dHW0}) is still the correct super-field equation to be solved up to quadratic order in derivatives and auxiliary fields \cite{Brizi:2009nn}. The latter approximation, however, also applies in the gravitational sector where, in the super-conformal approach, amounts in requiring that the $F$-term of the compensator field $\Phi$ is also suppressed: 
$F_{\Phi} \ll 1$. Modulo an overall rescaling, 
\be
F_\Phi = e^{K/2} \Big(\overline W + \frac 13 K_M F^M\Big) \,.
\ee
where  $F^M=g^{\bar M M} [\partial_{\bar M} \overline W +(\partial_{\bar M} K) \overline W]$. 
When $\langle W_0 \rangle \sim {\cal O}(\epsilon)$, or more generally $\exp(K/2) W \sim \epsilon$, 
$F_\Phi \sim \epsilon$ and cubic or higher auxiliary fields of the compensator can safely be neglected.
In this situation, the above analysis valid in global SUSY applies also in SUGRA with {\it no}
extra complication.

Notice that the freezing of a heavy field has to be performed at the level of $W$ and $K$ and not
of the component Lagrangian. Indeed, the structure of the SUSY lagrangian is such that several couplings appear to be proportional to the heavy field mass. In this case, the decoupling does not arise
since all the heavy fields interactions proportional to these couplings are unsuppressed and
have to be considered. As a matter of fact, such leading terms of the heavy field integration are automatically retained when freezing $H$ in $W$ and $K$. In order to explicitly see that, it might be useful to derive eq.(\ref{Vagree}) in a component formalism.
The full effective scalar potential, once the heavy moduli have been integrated out,  reads\footnote{For simplicity of notation, below and throughout the paper, we use the same notation to denote a chiral superfield and its lowest scalar
component, since it should be clear from the context to what we
are referring to.} 
\be
V_{full}= V(\langle H\rangle, L) + V_{int} (\langle H\rangle, L), 
\label{Vmicro}
\ee
where 
\be
V=e^K (F^M F_M - 3|W|^2)
\label{Vmicro2}
\ee
is the microscopical scalar potential and
\be
V_{int} = - \frac 12 V_IV^{IJ}V_J|_{H=\langle H\rangle} 
\label{deltaV2}
\ee
is the potential term induced by a gaussian integration of the  heavy moduli. Corrections to eq.(\ref{deltaV2}), due to cubic or higher terms in the heavy field fluctuations are negligible.  For simplicity, in eq.(\ref{deltaV2}) we have collected the holomorphic and antiholomorphic indices in a single notation, $I={i,\bar i}$, $V_I=\partial_I V$, and $V^{IJ}$ is the inverse matrix of $V_{IJ}=\partial_I\partial_JV$.  
It is not difficult to see that  $\partial_I V \sim {\cal O}(\epsilon)$, implying that 
the heavy field fluctuations $\hat H^i$ have to be explicitly integrated out, giving a correction of ${\cal O}(\epsilon^2)$ to the effective scalar potential.\footnote{One can neglect the heavy field fluctuations $\hat H^i$ even at the level of the scalar potential, if the 
light fields are kept dynamical in evaluating the vacuum shifts $H_1^i$. 
In other words, by computing $H_1^i(L)$, with $H^i = \langle H^i\rangle + \hat H^i = H_0^i+\epsilon H_1^i(\langle L\rangle +\hat L) + {\cal O}(\epsilon^2)+\hat H^i$. This was actually the procedure followed in \cite{LaI} to establish the equivalence of the full and simple theories.}
Given the form of eq.(\ref{deltaV2}), the knowledge of the potential at ${\cal O}(\epsilon^2)$ requires to compute the ${\cal O}(\epsilon)$ terms in $\partial_I V$ and the ${\cal O}(1)$ in $V^{IJ}$, the latter arising from the inverse of $V_{i\bar j}$. One gets
\bea
\partial_i \partial_{\bar j}  V|_0  & = & e^K \partial_{\bar j} \overline F_{\bar k,0} \, g^{\bar k l}\, \partial_i F_{l,0}
\,,\label{V02H} \\
\partial_i V|_1  & = & e^K \partial_i F_{l,0} F^l  \,,
\label{V0V1H}
\eea
where $F_{i,0}=\partial_i W_0$.
Using eqs.(\ref{V02H}) and (\ref{V0V1H}),  we find 
\be
V_{int} = -e^K  F^i \tilde g_{i\bar j}  \overline F^{\bar j}\,,
\label{deltaV}
\ee
where $\tilde g_{i\bar j}$ is the inverse of $g^{\bar j i}$, not to be confused with $g_{i \bar j }$.
The full effective scalar potential reads then 
 \bea
V_{full}  & = &  e^K \Big( g^{\bar M N} \overline  F_{\bar M} F_N - 3 |W|^2 \Big) -  e^K F^i \tilde g_{i\bar j}  \overline F^{\bar j}+{\cal O}(\epsilon^3)  \nn \\
& = & \epsilon^2  e^K\Big( \tilde g^{\bar \alpha \alpha} F_{\alpha} \overline F_{\bar \alpha} - 3 |W|^2\Big) +{\cal O}(\epsilon^3) = V_{sim}+{\cal O}(\epsilon^3)
\,, \label{VfullFnoY}
\eea
where $\tilde g^{\bar \alpha \alpha}$ is the inverse of $g_{\alpha \bar \alpha}$, the metric appearing in the simple model where the heavy fields are frozen, and the following matrix identity has been used:
\be
\tilde g^{\bar \alpha \alpha} = g^{\bar \alpha \alpha} - g^{\bar \alpha i} \tilde g_{i\bar j}g^{\bar j \alpha}\,.
\label{MatrixId}
\ee
Being the potential (\ref{VfullFnoY}) of ${\cal O}(\epsilon^2)$, it is enough  to keep the leading terms $H_0^i$ for the position of the VEV's in the heavy directions, finally  recovering eq.(\ref{Vagree}).

\subsection{${\cal O}(1)$ Yukawa couplings}

A necessary generalization of the set up in \cite{LaI} is the introduction of couplings of ${\cal O}(1)$ between charged fields, i.e., to allow  non-suppressed couplings in the light field sector. 
This extension forces us to distinguish, among the light fields $L$,  between the charged fields and the moduli. As is going to be clear below, it is also useful to distinguish between charged fields with ${\cal O}(\epsilon m_H)$ and ${\cal O}(m_H)$ VEV's.
We denote by $M^\mu$ ($\mu=1,\ldots,n_M$) the light moduli (including gauge singlets), by $Z^{\hat \alpha}$ ($\hat \alpha=1,\ldots,n_Z$)  the charged fields with ${\cal O}(m_H)$ VEV's and by $C^\alpha$ ($\alpha=1,\ldots,n_C$) the charged fields with ${\cal O}(\epsilon m_H)$ VEV's. We use calligraphic letters ${\cal A},{\cal B},\ldots$, to
collectively denote all the light field indices: ${\cal A}=(\alpha,\hat \alpha,\mu)$  and $M,N,\ldots$ to collect all fields indices, heavy and light, $M=(i,{\cal A})=(i,\alpha,\hat \alpha,\mu)$. Finally, we denote by $L^{\cal A}=(C^\alpha,Z^{\hat\alpha}, M^\mu)$ and by  $\phi^M=(H^i, C^\alpha,Z^{\hat\alpha}, M^\mu)$ the set of all light and of all light+heavy fields, respectively. The superpotential is taken as follows:
\be
W = W_0(H^i) +\eta\, \widetilde W_0(H^i,M^\mu,Z^{\hat \alpha},C^\alpha)  + \epsilon \, W_1(H^i,M^\mu,Z^{\hat \alpha},C^\alpha) \,.
\label{WYukC}
\ee
In eq.(\ref{WYukC}), $\widetilde W_0$ and $W_1$ are gauge-invariant polynomials in the charged fields $C^\alpha$, with field-dependent couplings: 
\bea
\widetilde W_0 & = &  Y_{3,\alpha\beta\gamma}(H^i,M^\mu,Z^{\hat\alpha}) C^\alpha C^\beta C^\gamma +{\cal O}(C^4) \,, \nn \\
W_1 & = &  \widetilde W_1(H^i,M^\mu, Z^{\hat \alpha})+ \mu_{2,\alpha\beta}(H^i,M^\mu,Z^{\hat \alpha}) C^\alpha C^\beta +{\cal O}(C^3)\,.
\label{WYukCX2}
\eea
The superpotentials $\widetilde W_0$ and $W_1$ do not need to admit a polynomial expansion in the $Z^{\hat\alpha}$ fields and can effectively be treated as moduli. The requirement that $W_0$ give a supersymmetric mass of ${\cal O}(m_H)$ to the heavy fields and that the fields $C$ have a mass of ${\cal O}(m_L)$
fix $W_0, W_1$ and $\mu_{2,\alpha\beta}$ to be of  ${\cal O}(m_H)$. 
The parameter $\eta$ is a dummy variable which will be useful in what follows, but that eventually will be taken to be equal to 1.  
The K\"ahler potential is of the form 
\be
K = K_0 +K_{1,\alpha\bar \beta} C^\alpha \bar C^{\bar \beta} + (K_{2,\alpha\beta} C^\alpha C^\beta
+c.c) + {\cal O}(C^3)\,, 
\label{KPot}
\ee
with $K_0$, $K_1$ and $K_2$ arbitrary functions of $H^i$, $M^\mu$ and $Z^{\hat\alpha}$ and their complex conjugates, constrained only by gauge invariance.

Our aim is to compare the theory defined above by $W$ and $K$ (the full theory) with the simple effective one where the $H^i$ are frozen at their leading VEV's $H_0^i$.
As before, we solve eq.(\ref{dHW0}) perturbatively in $\epsilon$ and,  at each order in $\epsilon$, 
further expand in $\eta$:
\be
H^i = H_0^i  + \eta \delta H_0^i(L) + {\cal O}(\eta^2) + \epsilon \Big[ H_1^i(L)  + {\cal O}(\eta) \Big]+\ldots 
\label{deltaHexp}
\ee
where, as usual, $H_0^i$ are defined by $\partial_i W_0(H_0) = 0$. 
The effective K\"ahler and superpotential read
 \bea
 W_{full} & = & W_{sim} +\eta^2 \Big(\frac 12 \partial_i \partial_j W_0 \delta H_0^i \delta H_0^j + \partial_i \widetilde W_0 \delta H_0^i\Big)+ 
  \epsilon \eta \Big(\partial_i \partial_j W_0 \delta H_0^i H_1^j + \partial_i \widetilde W_0  H_1^i\nn \\
 &&+  \partial_i W_1\delta H_0^i \Big)+  \epsilon^2 \Big(\frac 12\partial_i \partial_j W_0 H_1^i H_1^j  +\partial_i W_1H_1^i\Big) + {\cal O}(\eta^3,\eta^2 \epsilon,\eta \epsilon^2,\epsilon^3) \,,  \label{WeffSUSY1} \\
 K_{full} & = &  K_{sim} +\eta \Big[ \partial_i K_{sim} \delta H_0^{i}  + \partial_{\bar i} K_{sim} \delta \bar H_0^{\bar i} \Big] + {\cal O}(\epsilon, \eta^2)\,.
 \label{KeffSUSY1}
 \eea
The leading shift $\delta H_0^i$ is
 \be
 \delta H_0^i =  - W_0^{ij} \partial_j \widetilde W_0\,, 
\label{deltaHSUSY}
 \ee
with $W_0^{ij}\sim 1/m_H$ and $H_1^i$ as in eq.(\ref{H1SUSY}).
By plugging back eqs.(\ref{H1SUSY}) and (\ref{deltaHSUSY}) in eqs.(\ref{WeffSUSY1}) and (\ref{KeffSUSY1}), one easily finds
 \bea
 W_{full} & = & W_{sim} -\frac 12 \eta^2 \partial_i \widetilde W_0 W_0^{ij} \partial_j \widetilde W_0 - \epsilon \eta 
 \partial_i \widetilde W_0 W_0^{ij} \partial_j W_1 - \frac 12 \epsilon^2  \partial_i W_1 W_0^{ij} \partial_j W_1+\nn \\
 && {\cal O}(\eta^3,\eta^2 \epsilon,\eta \epsilon^2,\epsilon^3) \,,  \label{WeffSUSY} \\
 K_{full} & = &  K_{sim} -\eta \Big[\partial_i K_{sim} W_0^{ij}  \partial_j \widetilde W_0 + \partial_{\bar i} K_{sim} \overline W_0^{\bar i\bar j}  \partial_{\bar j} \widetilde{\overline W_0} \Big] + {\cal O}(\epsilon, \eta^2)\,.
 \label{KeffSUSY}
 \eea
Let us now see the structure of the new induced operators and their  possible relevance, recalling that  the meaningful region we can explore in the light charged field directions is defined to be $|C|/m_H\lesssim {\cal O}(\epsilon)$.  We will assume that $H^i\sim {\cal O}(1)$ to simplify the scaling analysis that follows.
The leading $C$-dependent terms of $W_{full}$ and $K_{full}$  (which are in $W_{sim}$ and $K_{sim}$) are of ${\cal O}(\epsilon^3 m_H^3)$ and
${\cal O}(\epsilon^2 m_H^2)$ respectively, and it is easily shown that correspondingly the leading $C$-dependent terms $V(C)$ of the scalar potential  are of ${\cal O}(\epsilon^4 m_H^4)$.  Notice that it is crucial to take $\exp(K/2) W_0 ={\cal O}(\epsilon m_H)$ in SUGRA, otherwise terms of ${\cal O}(\epsilon^3 m_H^4)$ would appear in the scalar potential, invalidating the equivalence of the simple and full theories. It is straightforward to see that all the induced couplings appearing in $W_{full}$ are at most of ${\cal O}(\epsilon^4 m_H^3)$ and those appearing in $K_{full}$ of ${\cal O}(\epsilon^3 m_H^2)$, 
so that 
\be
V(C)_{full} = V(C)_{sim}+{\cal O}(\epsilon^5 m_H^5)\,.
\label{Vc}
\ee
Eq.(\ref{Vc}) implies that the e.o.m. of the light fields $C^\alpha$ are the same in both approaches up to ${\cal O}(\epsilon^3)$, which is the first non-trivial order for these fields, being by assumption $\langle C^\alpha \rangle \sim {\cal O}(\epsilon m_H)$.  Of course, as far as the $C$--independent ${\cal O}(\epsilon^2)$ scalar potential is concerned, eq.(\ref{Vagree}) still applies, implying, in particular, that the leading e.o.m. of the fields $Z$ and $M$ are identical in the full and simple theories.
Notice that the ${\cal O}(\epsilon^5 m_H^5)$ terms in eq.(\ref{Vc}) arises only from ${\cal O}(\epsilon)$ corrections to coefficients of operators
present in $V(C)_{sim}$ and not from the new higher derivative operators induced by the heavy field integration. The latter, as we will see below, are sub-dominant, being at most of ${\cal O}(\epsilon^7 m_H^6)$.

The structure of the higher dimensional operators which are generated by the heavy field integration is easily seen from $W_{full}$ and $K_{full}$. The term proportional to $\eta^2$ in $W_{full}$  gives rise to new induced couplings between the charged fields of the form $Y_{N_i} Y_{N_j} C^{N_i+N_j}$. Their coefficients scale as $1/m_H$, which is higher than the natural scale ${\cal O}(1)$ for such operators. It is easy to see that the coefficients of all the higher dimensional operators induced by the terms in $W_{full}$ proportional to $\epsilon \eta$ and $\epsilon^2$ are smaller than their natural values.
Similarly, the terms proportional to $\eta$ in the K\"ahler potential $K_{full}$ give rise to $C^N$ higher dimensional operators with coefficients of ${\cal O}(1/m_H)$, higher than their ${\cal O}(1)$ natural values, implying that holomorphic cubic terms in $C$ are not reliable in the simple effective K\"ahler potential.
One can also compute the structure of the lowest dimensional induced operators appearing in the scalar potential $V_{full}(C)$, but  possibly not present in $V_{sim}(C)$. We get
\bea
\delta V(C)  \sim &&
\frac{Y_{N_i} Y_{N_j}Y_{N_k}}{m_H}  C^{N_i+N_j+N_k-2}+ \frac{Y_{N_i} Y_{N_j} \mu_{M_k} \epsilon}{m_H} C^{N_i+N_j+M_k-2} \nn \\
+&&  \frac{Y_{N_i}\mu_{M_j}\mu_{M_k} \epsilon^2}{m_H} C^{N_i+M_j+M_k-2}+\ldots \,.
\label{potential}
\eea
where we have schematically denoted by $Y_{N_i}$ and $\mu_{M_k}$ couplings and their derivatives with respect to $H$, $M$ and $Z$, we have omitted generic $M$ and $Z$ dependent coefficients and we have not distinguished between fields and their complex conjugates. 
The ellipsis contains terms which are of the same order or higher in $\epsilon$.
The lowest dimensional operators appearing in $\delta V(C)$ are of order $C^5$ but with suppressed couplings. The only operators with coefficients higher than their natural values are those appearing in the first term of eq.(\ref{potential}), of ${\cal O}(C^7)$. 
Notice that in the limit in which $Y_N$ are of ${\cal O}(\epsilon)$, the superpotential (\ref{WYukC}) is effectively of the form (\ref{Wint}) and the results of \cite{LaI} apply, namely $\delta V(C)={\cal O}(\epsilon^3)$. This explains why the minimum sum of powers of $Y_N$ and $\epsilon$ in eq.(\ref{potential}) is three.

It is straightforward to explicitly check that cubic or higher order terms in the light auxiliary fields (including the compensator in SUGRA) are all at most of the same order as the correction terms appearing in eq.(\ref{potential}), so that they can safely be ignored.
As a further check of this, we have also derived the scalar potential in an on-shell component approach, which amounts to sum all
powers of the light auxiliary fields, including the compensator field, and found complete agreement with eq.(\ref{potential}).
Needless to say,  even when the approximate manifestly SUSY equation $\partial_H W=0$ would give $\delta V(C)=0$, such as in the case when $Y_N$, $W_1$ and $\mu_M$ are all independent of $H$, higher order terms in the auxiliary fields would still give rise to 
terms of the form (\ref{potential}), if K\"ahler mixing terms between $H$ and the light fields are present.

\section{Chiral and vector multiplets}

We introduce in this section the vector multiplets by switching on the gauge couplings. 
We assume that at the vacuum a gauge group ${\cal G}$ is spontaneously
broken to a subgroup ${\cal H}$ at a scale parametrically larger than $\epsilon$.  The gauge group ${\cal H}$ might further be broken to a subgroup, but only at scales of ${\cal O}(\epsilon m_H)$. We denote by $X_A^M$ and $\overline X^{\bar M}_A$ the holomorphic and anti-holomorphic Killing vectors generating the (gauged) isometry group ${\cal G}$, defined as $\delta \phi^M = \lambda^A X^M_A$, $\delta \bar\phi^{\bar M} = \lambda^A \overline X^{\bar M}_A$, with $\lambda^A$ infinitesimal real parameters. The corresponding $D$-terms are
\be
D_A = i  X^M_A G_M = - i \overline X^{\bar M}_A  \overline G_{\bar M}\,,
\label{DFrel}
\ee
where $G = K + \log |W|^2$ and  $G_M = \partial_M G = F_M/W$ are the K\"ahler invariant function 
and its derivatives and $F_M=D_M W = \partial_M W +(\partial_M K) W$
is the K\"ahler covariant derivative.\footnote{We have not included constant Fayet-Iliopoulos terms $\xi_A$ in eq.(\ref{DFrel}) motivated by the fact that they do not seem to appear in string derived models of any sort. In SUGRA, their presence requires superpotentials which are gauge-invariant only modulo an $R$--symmetry shift, namely $X^i_A \partial_i W = \xi_A W$ for $U(1)$ transformations.} 
We denote by capital latin letters $A,B,\ldots=1,\ldots, \, {\rm adj}( {\cal G})$, the gauge group indices, not to be confused  with the light field indices ${\cal A},{\cal B},\ldots$ introduced before.  For simplicity of presentation, we take the holomorphic gauge kinetic functions $f_{AB}$ diagonal in the gauge indices, so that 
\be
f_{AB} = \delta_{AB} f_A(H^i,M^\mu,Z^{\hat \alpha}, C^\alpha)
\ee
are generic holomorphic functions and $ {\rm Re}\, f_A =1/g_A^2$, with $g_A$ the  coupling constants of ${\cal  G}$. 

Contrary to the the pure $F$-term case, the condition $\partial_i W_0=0$ may not fix all the fields $H^i$.  Indeed, gauge invariance relates the derivatives of $W_0$ by
\be
X^i_{A}
\partial_i W_0 = 0\,,
\label{XiW0}
\ee
so that in general they are not linearly independent and some of the fields may remain unfixed. 
Thus, if the $X^i_A$ are not all vanishing, one can always choose a basis in which some of the fields do not appear at all in $W_0$. It is very simple to explicitly construct such a  basis  for the relevant case where the $H^i$ are moduli shifting under a $U(1)$ gauge symmetry.
If $\delta H^i = i \xi^i \Lambda$,  with $\Lambda$ the chiral super-field associated to the $U(1)$ transformation, one can always choose to parametrize $W_0$ in terms of, say, $M^1\equiv H^1$ (with $\xi^1\neq 0$)  and the $n_H - 1$ gauge invariant operators $H_{GI}^{i}=\xi^1 H^{i}-\xi^i H^1$, with $i>1$.
One can invert this relation and use the fields $H^i_{GI}$ in $W_0$. In this field basis, the only non-vanishing isometry component is $X^1$. From eq.(\ref{XiW0}) we immediately see that $W_0$  is independent of $M^1$, depending only on the $n_H-1$ gauge invariant fields $H_{GI}^i$. Of course the same argument can be repeated for each $U(1)$ generator independently.
Hence there is no well defined meaning in freezing $M^1$. This is not even a meaningful gauge invariant statement, since $\delta M^1 \neq 0$ and thus $M^1$ will necessarily enter in $W$ (if any) in a gauge invariant combination with light fields.
One might get rid of $M^1$ by gauging it away, namely choosing a gauge (which is not the Wess-Zumino gauge) where it is a constant. This is possible, but then its dynamics will reappear in the $U(1)$ vector super field, more precisely in 
the longitudinal component of the gauge field and the lowest auxiliary component  of the  $U(1)$ vector super field, which is now dynamical. If the vector field is sufficiently massive and is integrated out, the effects of $M^1$ will eventually appear in new contributions to the K\"ahler potential of the remaining light fields. 
A direct relevant consequence of the above result is the impossibility of naively neglecting moduli fields responsible for field-dependent Fayet-Iliopoulos terms and then forgetting the implicit gauge-fixing taken behind this choice, as sometimes done in the literature in local string constructions, where all moduli dynamics is neglected altogether. Clearly, this obstruction has nothing to do with gravity and is purely dictated by gauge invariance. This observation actually dates back to \cite{ArkaniHamed:1998nu} and was more generally expressed in \cite{Binetruy:2004hh}.  In conclusion, for all proper heavy fields we must impose 
\be
X^i_A= 0\,.
\label{Xi0}
\ee
In addition to eq.(\ref{Xi0}), for simplicity we also assume that the remaining isometry components do not depend on $H^i$,
$X_A^{\cal A} = X_A^{\cal A}(L^{\cal B})$.

Since vector multiplets do not appear in $W$, the manifestly SUSY integration of chiral multiplets is not naively affected at leading order by their presence. However, even at the level of two derivatives (quadratic auxiliary fields), terms with covariant derivatives in the K\"ahler potential may now appear, coming from terms of the form $D F$ or $D^2$.\footnote{We thank the authors of 
\cite{Brizi:2009nn} for discussions on this issue and sharing with us a preliminary draft of their paper.}  However, these terms are necessarily further suppressed by powers of $1/m_H$. This is easily seen by noticing that a $D$-term is the $\theta^2 \bar \theta^2$ component of a vector multiplet
as opposed to an $F$ term of a chiral field,  which is its $\theta^2$ component. This implies that
the $DF$ and $D^2$ generated terms in $K$ arises from terms with at least two and four covariant derivatives, respectively. They are then necessarily suppressed by $\epsilon$ or $\epsilon^2$ with respect to the $F^2$ generated terms and can thus be neglected.

When the gauge group ${\cal G}$ is unbroken at ${\cal O}(1)$, namely when there are no $Z$ fields and all the moduli $M$ are neutral, the effective K\"ahler and superpotential are as in eqs.(\ref{WeffSUSY}) and (\ref{KeffSUSY}). By plugging eqs.(\ref{H1SUSY}) and (\ref{deltaHSUSY}) in $f_{AB}$ we get the effective holomorphic gauge kinetic functions
\be
f_{AB,full} = f_{AB,sim} + \partial_i f_{AB} \delta H_0^i + {\cal O}(\epsilon, \eta^2)\,.
\label{fab}
\ee
Eq.(\ref{fab}) implies that all $C$--independent terms and $C$--dependent ones up to $C^{N-1}$ (included) entering in the $f_{AB}$ are reliable at ${\cal O}(1)$ in $\epsilon$. The effective K\"ahler potential (\ref{KeffSUSY}) gives rise to corrections to the D-terms  of the schematic form
\bea
D_{A,full} -D_{A,sim} &  = &   i X_A^{\alpha} \partial_\alpha  \Big[ \partial_i K_{sim}(\eta \delta H_0^{i}+\epsilon H_1^i)  + \partial_{\bar i} K_{sim} (\eta \delta \bar H_0^{\bar i}+\epsilon \bar H_1^i) \Big] + {\cal O}(\epsilon^2, \eta^2,\epsilon\eta)\nn \\
&& = \frac{Y_{N_i}}{m_H} C^{N_i+2}  + \epsilon C^2+ \frac{\epsilon \mu_{M_i}}{m_H} C^{M_i+2} +\ldots \,, 
\label{DeffSUSY}
\eea
where  we have set $\eta=1$ in the second row of eq.(\ref{DeffSUSY}) and where, for concreteness, 
we have counted the powers of $C$ (including their complex conjugates) by taking linear realizations of the gauge group ($X_A^\alpha$  proportional to $C^\alpha$). Gauge invariance of $\widetilde W_0$ and $W_1$ has been used, constraining  $X_A^\alpha \partial_\alpha \delta H_0^i$ and $X_A^\alpha \partial_\alpha H_1^i$ to vanish. 

We can easily extract from eq.(\ref{DeffSUSY}) the lower dimensional operators generated by the heavy field integration, appearing in the $D$--term scalar potential $V_{D,full}$ but not present in $V_{D,sim}$. 
They are of the form $g^2 Y_{N_i} C^{N_i+4}/m_H$ and $\epsilon \mu_{M_i}g^2 C^{M_i+4}/m_H$. 
Both are at most of ${\cal O}(\epsilon^7 m_H^6)$ and hence irrelevant.
As already mentioned, in presence of vector multiplets, quadratic terms in the auxiliary fields of the form
$DF$ and $D^2$ are missed. It is useful to explicitly see how this discrepancy arises in our class of models.  By construction, all the effective operators one obtains from the effective $D$-term scalar potential by a manifestly SUSY integration, where covariant derivatives are neglected, give rise to only operators proportional to $g^2$, as above. On the other hand, it is obvious that in presence of
a K\"ahler mixing term between the charged fields and the heavy fields, higher order operators with coefficients proportional  to $(g^2)^2/m_H^2$ should be expected. By explicitly computing the scalar potential in components, indeed, we find
\be
\delta V_D \supset   \frac{g^2 Y_{N_i}}{m_H} C^{N_i+4}+\frac{\epsilon g^2 \mu_{M_i}}{m_H}C^{M_i+4}+ 
\frac{(g^2)^2}{m_H^2} C^8\,.
\label{deltaVD}
\ee
The first two terms are exactly those found in the manifestly SUSY case, whereas the latter arises
from the undetected $D^2$ terms and, as expected, is negligible, being of ${\cal O}(\epsilon^8 m_H^6)$.

In presence of $Z$ fields and non-neutral light moduli, the gauge group is spontaneously broken and the analysis is  more involved, because extra states become massive, the real scalar fields associated to the would-be Goldstone bosons eaten by the heavy gauge fields $A_{\hat a}$ and the gauge fields $A_{\hat a}$ themselves, where  we have splitted  the gauge index $A=(a,\hat a)$, $\hat a\in {\cal G/H}$,  $a\in {\cal H}$.  Such fields cannot be frozen to their VEV's but they should be properly integrated out. 
A manifestly SUSY integration of a vector super field requires that $\langle D \rangle /m_V^2 \ll 1$, 
with $m_V$ the gauge field mass.  In our case this condition is always satisfied, as shown in Appendix A, being $\langle D_{\hat a}\rangle  ={\cal O}(\epsilon^2)$.
A vector super field is supersymmetrically integrated out, neglecting covariant derivative terms coming from the holomorphic gauge field action, by setting \cite{ArkaniHamed:1998nu}
\be
\partial_{V_{\hat a}} K = 0\,,
\label{KVeq0}
\ee
$V_{\hat a}$ being the vector-superfields associated to the broken generators.
For typical charged field K\"ahler potentials, eq.(\ref{KVeq0}) does not admit a simple constant solution, so that it is in general hopeless to freeze a heavy vector super field. Integrating out a vector super field implies a choice of gauge fixing. The physical gauge where one gets rid of the eaten Goldstone bosons and their superpartners is the super-field version of the unitary gauge. On the other hand, it is practically easier to work in a gauge where a chiral field $Z$ (or a non-neutral modulus $M$) is frozen at its VEV $Z_0$ (or $M_0$). The field choice is arbitrary, provided it has a non-vanishing component in the would-be Goldstone direction. This can be done for each broken generator so that ${\rm dim} \,{\cal G}/{\cal H}$ light chiral multiplets  (or combinations thereof) are gauged away from the theory.
Let us denote by $L^{{\cal A}^\prime}$ the remaining directions, with ${\cal A}^\prime = 1,\ldots, n_L - {\rm dim} \,{\cal G}/{\cal H}$, and by 
$V_{\hat a}^{0}$ the solution to eq.(\ref{KVeq0}). By plugging back in the Lagrangian $V_{\hat a}=V_{\hat a}^{0}$, we get the SUSY effective theory
with heavy vector fields integrated out. As far as the chiral fields are concerned, the correction terms one obtains from the holomorphic gauge kinetic terms 
are negligible, being suppressed by four covariant derivatives with respect to the corrections terms coming from the effective K\"ahler potential 
$K^\prime = K(V_{\hat a}=V_{\hat a}^{0})$.\footnote{If all the components of $V_{\hat a}^{0}$ are much smaller than the heavy vector mass to the appropriate power, an explicit expression for $K^\prime$ can be given by expanding the vector-super fields $V_{\hat a}$ up to quadratic order, in which case one has
\be
V_{\hat a}^0 = -  \widetilde m_{\hat a\hat b}^{-2} K_{\hat b}/2 \,,
\ee
where $\widetilde m^2_{\hat a\hat b}= 2 \langle g_{M\bar N} X_{\hat a}^M   \overline X^{\bar N}_{\hat b}\rangle  $ is the non-canonically normalized mass matrix for the gauge fields and $K_{\hat a} = \partial_{V_{\hat a}} K|_{V=0}$.
Plugging back in $K$, it gives
\be
K^\prime = K(V_{\hat a}=0) - K_{\hat a} \widetilde m_{\hat a\hat b}^{-2} K_{\hat b}/4\,.
\label{KSUG2}
\ee}
The gauge fixing should also be plugged in $W$, giving rise to a superpotential $W^\prime$, which is a function of the $H^i$ and $L^{{\cal A}^\prime}$. The effective $D_a^\prime$ term now reads
\be
D_a^\prime = i X^{{\cal A}^\prime}_a G_{{\cal A}^\prime}^\prime  = -i \overline X^{\bar{\cal A}^\prime}_a   G_{\bar{\cal A}^\prime}^\prime\,,
\label{DSUG}
\ee
with $X^{{\cal A}^\prime}_a$ the components of the pulled back isometry vectors on the K\"ahler invariant function $G^\prime$, defined by the gauge fixing. Since the latter is linear in the fields, the pull-back is trivial and $X^{\alpha\prime}_a$ are nothing else than the original isometry vector components along the non-gauged away directions $L^{{\cal A}^\prime}$. After having integrated out the massive vector fields and their scalar partners,  we get an intermediate effective theory given by the K\"ahler potential $K^\prime$, isometries
$X_a^{{\cal A}^\prime}$, superpotential $W^\prime$ and gauge kinetic functions $f_a$, with ${\rm Re}\, f_a = 1/g_a^2$. 
No field with non-vanishing VEV and charged under ${\cal H}$ appears in this theory. All the $Z$ and the non-neutral $M$ have been either gauged away or appear as gauge singlet combinations in the intermediate theory,
effectively behaving as new gauge invariant fields. 
By expanding $K^\prime$ in powers of the charged fields we get a K\"ahler potential of the form (\ref{KPot}) (where $K_0^\prime$, $K_1^\prime$ and $K_2^\prime$, in turn,
can be expanded in powers of the heavy vector mass) and we are effectively back to the case discussed before of unbroken gauge group.
We can now integrate out the $H$ and
compare the resulting theory with  the simple effective theory $K^\prime_{sim} = K^\prime(H_0)$, $W^\prime_{sim} = W^\prime(H_0)$, $f_{a,sim}=f_a(H_0)$.  As expected,  $K^\prime_{sim} $ and $W^\prime_{sim}$ coincide with the K\"ahler and superpotential terms obtained by 
integrating out the heavy vector fields in the simple theory, when the same gauge-fixing taken in the full theory is used.

\section{Conclusions}

We have shown that in a wide class of SUSY theories, including vector fields and ${\cal O}(1)$ Yukawa couplings for charged fields,  it is reasonable to neglect the dynamics of heavy chiral fields and yet retain a predictive simple effective field theory at low-energies. It is enough that the heavy fields sit at an approximately SUSY vacuum in their directions, even when SUSY is completely broken in the light field directions, and (in SUGRA) that $W$ is sufficiently small so that curvature corrections and  SUSY breaking mass splitting are much smaller than the heavy field masses.
In our class of models, where a chiral field is heavy through a large superpotential term, by construction only gauge-invariant fields can be frozen. No constraint on the K\"ahler potential is assumed, provided that it is sufficiently regular.  
It is not difficult to extend the results of our paper to the situation where more mass hierarchies are present. As expected, new expansion parameters $\epsilon_i \equiv m_i/m_H$, where $m_i$ are the new mass scales, appear and all the analysis still applies, being only more cumbersome. 

One might wonder whether light moduli, with mass of ${\cal O}(m_L)$, can somehow also be frozen. The generic answer is clearly no, but it is interesting to see under what special conditions this may be done. 
The first obvious requirement to freeze the light moduli $M$ is that $F^M\simeq 0$. This is a 
non-trivial condition since the backreaction induced by other SUSY breaking effects on them is no longer suppressed by a heavy mass. If it does not occur, not only the light moduli cannot be neglected, but
it is generally also hard to integrate them out and the resulting effective theory cannot be described by a SUSY theory.
The best one can hope is that their interactions are negligible and just retain the effect of their $F$--terms, in which case one can compute the associated soft terms \cite{Kaplunovsky:1993rd}. Even if $F^M\simeq 0$, the condition $F_\Phi\ll m_L$ will not in general
be satisfied, so that a description of the resulting effective theory in terms of a standard two-derivative SUGRA action will not be available. On the other hand, if  $F^M\simeq 0$ and $F_\Phi\ll m_L$,  the analysis of our paper can be repeated almost step by step. In the reasonable hypothesis that the Yukawa couplings $Y$ do not depend on $M$, because, say,  the same symmetry forbidding an ${\cal O}(1)$ superpotential term $W_0$ for them might also forbid $M$--dependent ${\cal O}(1)$ Yukawa terms, one can just replace $W_0$ with $W_1$ and the couplings $Y_N$ with $\mu_M$. 
In this set-up, the theory that so far we have called ``simple" becomes the underlying theory and we call ``super-simple'' the resulting theory with the light moduli frozen.  It is straightforward to see, properly adapting eqs.(\ref{WeffSUSY}) and (\ref{KeffSUSY}) to this situation, that the induced couplings appearing in $W_{full}$ are at least of ${\cal O}(\epsilon^5 m_H^3)$, but those appearing in $K_{full}$ are now of ${\cal O}(\epsilon^2 m_H^2)$, since $C^2$ terms in $K$ may be induced with unsuppressed coefficients. The latter can be relevant  for the computation of holomorphic
soft mass terms. Hence, the super-simple theory can generally be reliable only if $F^M\simeq 0$, $F_\Phi\ll m_L$ and the induced $C^2$ terms in $K$ are somehow suppressed.

\section*{Acknowledgments}

We would like to thank Riccardo Rattazzi and Claudio A. Scrucca for useful discussions.

\appendix

\section{A detailed analysis of the vacuum}

In this appendix, along the lines of \cite{LaI},  we show in some more detail how the e.o.m. of the light scalar fields agree at leading order in the full and simple models in presence of vector fields. 
Since non-trivial e.o.m. for the fields $C$ appear only at ${\cal O}(\epsilon^3)$,
for simplicity we set them to zero, which is always a solution to their e.o.m., and only study the e.o.m.
for the remaining $M$ and $Z$ fields. In order to keep the notation as simple as possible, we omit in this appendix the subscript ``full'', being understood that any quantity with no specification arises in the full theory.

The new ingredient with respect to the analysis performed in \cite{LaI} is the $D$-term scalar potential
\be
V_D=\frac12 g_A^2 D_A^2\,.
\ee
We study the location of the vacuum in both theories in a series expansion in $\epsilon$:
\be
\langle \phi^M\rangle=\phi_0^M+\epsilon \phi_1^M+\epsilon^2 \phi_2^M+\ldots\,.
\ee
Although the $D$-term potential does not admit an expansion in $\epsilon$, being governed by generally ${\cal O}(1)$ gauge couplings,  at the vacuum the $D$-terms are related to $F$-terms and hence an expansion in $\epsilon$ is still possible.

At ${\cal O}(\epsilon^0)$, $\partial_i W_0(H_0) =0$  solve the $F$-term e.o.m. for the heavy fields and trivialize the corresponding ones for the light fields: $(\partial_i V_F)_0 = (\partial_{\cal A} V_F)_0 = 0$. Of course, due to the presence of $V_D$, this is no longer a sufficient condition but it is still necessary. In this way, the leading order VEV's for the heavy fields are fixed. 
At ${\cal O}(\epsilon^0)$ and $H^i=H_0^i$, the e.o.m. for the light fields are entirely given by the $D$-term potential:
\be
(\partial_{\cal A} V_D)_0 = \frac 12 (\partial_{\cal A} g_A^2) D_A^2 + g_A^2 D_A \partial_{\cal A} D_A = 0\,,
\label{eqD0}
\ee
evaluated at $\phi_0^M$, which admit the simple solution
\be
D_A(\phi_0^M) = 0\,.
\label{DA0}
\ee
When the gauge symmetry is unbroken, eq.(\ref{DA0}) is a solution to all orders in $\epsilon$.
Indeed, from eq.\eqref{DFrel} it is straightforward to deduce the following general bound,
\be
\sqrt{2} g_A D_A\le \sqrt{g^{\bar N  N}K_N K_{\bar N}} m_{AA}\,,
\ee
with $m_{AA}$ being the diagonal components of the gauge field mass matrix
\be
m^2_{AB} = 2 g_A\, g_B\, g_{M\bar N} X_A^M   \overline X^{\bar N}_B \,.
\label{Mgauge}
\ee
For spontaneously broken symmetries, another relation between $F$ and $D$--terms is valid at the vacuum, of the form $\langle D\rangle\sim \langle F^2 \rangle/m^2_V$ (see eq.\eqref{Dhatdyna} below), where $m_V$ is the typical scale of the heavy vector fields, parametrically larger than $\epsilon$.
Requiring the $F$-terms to be all at most of ${\cal O}(\epsilon)$, we conclude that at the vacuum $\langle D\rangle \lesssim {\cal O}(\epsilon^2)$. Eq.(\ref{DA0}) is the only sensible solution to eq.(\ref{eqD0}) for vacua with no ${\cal O}(1)$  SUSY breaking. Eq.(\ref{DA0}) also ensures that at ${\cal O}(\epsilon^0)$ the e.o.m. of the heavy fields are automatically satisfied at $H_0$, since $(\partial_i V_D)_0 = 0$. 

At ${\cal O}(\epsilon)$, the e.o.m. for the light fields are still given by the $D$-term potential only, so that
\be
(\partial_{\cal A}V_D)_1 = g_A^2 \Big[\partial_N D_A(\phi_0) \phi_1^N + \partial_{\bar N} D_A(\phi_0) \phi_1^{\bar N}\Big]\partial_{\cal A} D_A(\phi_0)=0\,.
\label{D1eq}
\ee
Two possible solutions can be taken. Either $\partial_{\cal A} D_A(\phi_0)=0$, which  implies $X^{\cal A}_A(\phi_0^M)= 0$, being $X^i_A = 0$,  
or $\partial_N D_A(\phi_0) \phi_1^N + \partial_{\bar N} D_A(\phi_0) \phi_1^{\bar N}=0$. The two situations correspond respectively to unbroken and broken
generators. Indeed, splitting the gauge index $A=(a,\hat a)$ in eq.(\ref{Mgauge}), with $a\in {\cal H}$, $\hat a\in {\cal G/H}$, we have
 \be
m^2_{\hat a\hat b} = {\cal O}(m_V^2), \ \ \ m^2_{\hat a a} = m^2_{a \hat a} =0,  \ \ \  m^2_{ab}=0\,.
\label{MgaugeSca}
\ee
By taking $g_A^2$ and the K\"ahler metric parametrically of order one, eq.(\ref{MgaugeSca}) gives
\be
\langle X_a^M\rangle  =0, \ \ \  \langle X_{\hat a}^M\rangle = {\cal O}(m_V)\,.
\label{Xsca}
\ee
So, we have
\bea
X_{M,a}(\phi_0) =  X^M_a(\phi_0) &  = &  0,  \ \ \ \ a\in {\cal H} \,, \nn \\
 \partial_N D_{\hat a}(\phi_0) \phi_1^N + \partial_{\bar N} D_{\hat a}(\phi_0) \phi_1^{\bar N} & = & 0, \ \ \ \hat a\in {\cal G/H}\,.
\label{eqD1L}
\eea
Eqs.(\ref{eqD1L}) imply that both $D_a$ and $D_{\hat a}$ vanish at ${\cal O}(\epsilon)$, in agreement with our previous argument that  $D_A \leq {\cal O}(\epsilon^2)$.
When eq.(\ref{eqD1L}) is satisfied, the e.o.m. for the heavy fields at ${\cal O}(\epsilon)$ are given by $V_F$ only and
fix $H_1^i$ as in the pure $F$-term case studied in \cite{LaI}, 
\be
H_1^i = - (\hat K^{-1})^i_{\bar j} (\overline G^{\bar j})_0\,,
\label{PhiH1sugra}
\ee
with $\hat K^{\bar i}_j  = g^{\bar i k}(\partial_k  G_j)_{-1}$ and $(\overline G^{\bar j})_0 = (\overline G_{\bar M})_0 g^{\bar M i}$, notation like in \cite{LaI}, evaluated at $H^i=H_0^i$. At the shifted vacuum $H_0^i+\epsilon H_1^i$, we have
\be
G_i  =  {\cal O}(1)\,,  \hspace{0.2cm}  G_{\cal A} = {\cal O}(1)\,, \hspace{0.2cm} G^i =   {\cal O}(\epsilon)\,, 
\hspace{0.2cm} G^{\cal A}  = {\cal O}(1)\,, \label{GdG} 
\ee
showing that the matching of the $F$-term part of the e.o.m. of the light fields at ${\cal O}(\epsilon^2)$, $(\partial_{\cal A} V_F)_{full} = (\partial_{\cal A} V_F)_{sim} + {\cal O}(\epsilon^3)$, continues to hold in presence of $D$-terms. 
Hence we can just focus on $V_D$. The term $(\partial_{\cal A}  V_D)_2$ can be written as follows:
\be
(\partial_{\cal A} V_D)_2  = g_{\hat a}^2(\phi_0) D_{\hat a}(\phi_0+\epsilon \phi_1+\epsilon^2 \phi_2) \partial_{\cal A}  D_{\hat a}(\phi_0)\,,
\label{D2eq}
\ee
where we have explicitly written the order at which the various quantities should in principle be known.

A similar expansion of the e.o.m. can be performed in the simple effective theory, where the $D$--term reads $ V_{D,sim} = g_A^2 (D_A)_{sim}^2/2$, with $(D_A)_{sim} = i X^{\cal A} _A \partial_{\cal A}  K_{sim}$, and all quantities evaluated at the leading frozen vacuum $H_0^i$.
At ${\cal O}(\epsilon^0)$ and ${\cal O}(\epsilon)$ we get
\be
(D_A)_{sim}(\phi_0)  =   0 , \ \ \
 (X^{\cal A} _a)_{sim}(\phi_0)   =   0 , \ \ \
 \partial_{\cal A}  D_{\hat a,sim}(\phi_0) \phi_{1,sim}^{\cal A}  + \partial_{\bar{\cal A}}  D_{\hat a,sim}(\phi_0) \phi_{1,sim}^{\bar{ \cal A}}  =  0\,,
\label{eq2D1L}
\ee
which implies that $D_{sim}(\phi_0) \sim {\cal O}(\epsilon^2)$.
The form of $(\partial_{\cal A}  V_D)_{sim}$ at ${\cal O}(\epsilon^2)$ is  the same as eq.(\ref{D2eq}), but written in terms of $(D_A)_{sim}$ and evaluated at $H_0^i$. Thus, the equivalence of the two descriptions requires that the following non-trivial relation holds:
\be
D_{\hat a}(\langle \phi^M  \rangle)  =   (D_{\hat a})_{sim}(H_0^i,\langle L^{\cal A}  \rangle) + {\cal O}(\epsilon^3)\,,
\label{DD}
\ee
with $\langle \phi \rangle$ and $\langle L \rangle$  expanded up to ${\cal O}(\epsilon^2)$. 
Luckily enough, we do not need to work out the vacuum up to ${\cal O}(\epsilon^2)$, since we can trade ${\rm dim}\, G$ e.o.m.
to write ${\dim }\, G$ relations between the $F$ and $D$--terms at the vacuum.\footnote{Notice that, modulo global gauge transformations,  the vacuum is uniquely determined since the ``missing'' ${\rm dim}\,G$ e.o.m. are provided by the ${\rm dim}\,G$ D-term constraints (\ref{DA0}).}  Indeed, by taking 
the combination of the e.o.m. ${\rm Im} (X_A^M \partial_M V) = 0$ (the real part being identically vanishing by gauge invariance:
$\delta_\lambda V = \lambda^A {\rm Re}( X_A^M \partial_M V)=0$), one easily derive the following equation (see e.g.\cite{Kawamura:1996wn,GomezReino:2007qi}:
\be
q_{AM\bar M} F^M \overline F^{\bar M} - \frac 12 D_B \Big[m_{AB}^2 + \delta _{AB} (F^M F_M - m_{3/2}^2)\Big] =0
\,. \label{DDdyna}
\ee
where $q_{A M\bar M} = \nabla_M \nabla_{\bar M} D_A$ and $m_{3/2}=\exp(K/2)W$.\footnote{If the gauge kinetic functions are not gauge-invariant (e.g. as required by anomaly cancellation), an extra term appears in eq.(\ref{Dhatdyna}). Being of ${\cal O}(D^2)\sim {\cal O}(\epsilon^4)$, it is completely negligible for our purposes. See e.g. \cite{GomezReino:2007qi} for a more general formula including these terms.} 
When $A=a$, the second term in eq.(\ref{DDdyna}) vanishes and the equations boil down to a set of constraints for the $F$-terms at the vacuum, dictated by gauge invariance.  When $A=\hat a$, instead, we can invert eq.(\ref{DDdyna}) to solve for $D_{\hat a}$:
\be
D_{\hat a} = 2m_{\hat a\hat b}^{-2} q_{\hat b M\bar M}F^M \overline F^{\bar M}+{\cal O}(\epsilon^4)\,.
\label{Dhatdyna}
\ee
A similar relation occurs in the simple effective theory upon replacing the indices $M$ and $\bar M$ with ${\cal A}$
and ${\bar{\cal A} }$. As expected, eq.(\ref{Dhatdyna}) tells us that $D_{\hat a} \sim {\cal O}(\epsilon^2)$, since the $F$--terms are all at most of ${\cal O}(\epsilon)$. In addition, since $F^i(\langle \phi^M \rangle )\sim {\cal O}(\epsilon^2)$, we see that
eq.(\ref{DD})  is easily proved:
\be
D_{\hat a}(\langle \phi^M \rangle) = 2m_{\hat a\hat b}^{-2} q_{\hat b \cal A{\bar{\cal A}}}F^{\cal A} \overline F^{\bar{\cal A}} (\phi_0^M ) +{\cal O}(\epsilon^3)
= (D_{sim})_{\hat a}(H_0^i,L_0^{\cal A}) +{\cal O}(\epsilon^3)\,.
\label{DDproof}
\ee
We have then established that even in presence of $D$--terms the location of the vacuum as computed by the simple effective theory is reliable.

 \section{A Numerical Curved Space Model with $U(1)$ Gauge Symmetry}

In this appendix we apply the results of the paper to a string-inspired SUGRA toy model with $n_H=2$, $n_M=1$, $n_Z=2$ and $n_C=0$,
where a Fayet-like  SUSY breaking mechanism occurs (see \cite{Gallego:2008sv,Dudas:2008qf} where, building on \cite{Dudas:2007nz},  
effective models of this sort have been analyzed in some detail).
Although the model contains just five complex fields, instead of hundreds as in realistic string models,
it is already sufficiently complicated to make an analytical study a formidable task. For this reason we opt here
for a numerical analysis. In particular we will show how the vacuum and the scalar mass spectrum as given by the simple effective model is in agreement with the full--fledged analysis. The K\"ahler and superpotential terms are taken as follows:
\bea
\hspace{-0.9cm} K & = & - 2 \log\Big[ (T+\bar T-\delta V_X)^{3/2} + \xi(S+\bar S)^{3/2}\Big]-\log(Z+\bar Z)-\log(S+\bar S) \nn \\
&& + \frac{\bar \phi \, e^{-2V_X} \phi}{(Z+\bar Z)^{n_\phi}}
+ \frac{\bar \chi\,  e^{2V_X} \chi}{(Z+\bar Z)^{n_\chi}}  \,, \label{K4sugra} \\
\hspace{-0.9cm} W & = & a Z^2 + b Z + S (c Z^2 + d Z + e) + m Z \phi \chi + \beta Z^2 \phi^{\alpha\delta/2} e^{-\alpha T} \,. \label{W4sugra}
\eea
The heavy fields $S$ and $Z$ mimic respectively the dilaton and a complex structure modulus of some IIB flux Calabi-Yau compactification, $T$ represents the overall universal light K\"ahler modulus, $\phi$ and $\chi$ are two charged fields with ${\cal O}(1)$ VEV's and opposite $U(1)$ charge
with respect to a $U(1)$ gauge field $A_X$. 
The holomorphic gauge kinetic function associated to $U(1)_X$ is taken to be $f_X = T$.
The K\"ahler modulus $T$ is not gauge invariant, but transforms as $\delta T = i \delta/2 \Lambda$, where $\delta \phi = i \Lambda \phi$,
$\delta \chi = -i \Lambda \chi$. The fields $S$ and $Z$ are gauge invariant.  We refer the reader to the appendix B of \cite{LaI}, where a similar superpotential was considered,  for an explanation of the various terms appearing in $W$ and the choice of parameters that follows.
The $U(1)_X$ $D$--term is equal to
\be
D_X = i X^M G_M = i X^M \partial_M K = \frac{|\chi|^2}{(2Z_r)^{n_\chi}} -\frac{|\phi|^2}{(2Z_r)^{n_\phi}} + \frac{3\delta T_r^{1/2}}{4(T_r^{3/2}+\xi S_r^{3/2})}\,,
\label{DXapp}
\ee
where the subscript $r$ denotes the real part of the field. The essential dynamics of the system is driven by the $D_X$ term. In order to
minimize the energy carried by the effective field-dependent FI term, the third term in eq.(\ref{DXapp}), $\phi$ acquires a non-vanishing VEV.
The mass term in the superpotential becomes then effectively a Polonyi-like term for the field $\chi$, which breaks SUSY.
We have taken\footnote{Notice that there is  a typo in the values of $a$ and $b$ in eq.(B.3) of \cite{LaI}, the correct values being the ones reported in eq.(\ref{explPARA}).} 
\bea
a &  = &  -2.55-10^{-13}\,, \ \ \ b = 25.5+2\cdot 10^{-12}\,, \ \ \  c = 0.25\,, \ \ \ d = -2.45\,, \ \ \ e=-0.5 \nn \\
\alpha& = & 1\,, \ \ \beta = -0.5\,,  \ \ \delta = 1\,, \ \ \ m = 2.44\times 10^{-12}\,, \ \ n_\chi = 1\,, \ \ n_\phi = 0\,, \ \ \ \xi =  0.1\,.
\label{explPARA}
\eea
The smallness of $m$, required to get a dS vacuum with a sufficiently small cosmological constant, justifies the location of the $Z\phi\chi$ operator in $W_1$. The SUSY VEV's of $S$ and $Z$, as given by $\partial_S W_0 = \partial_Z W_0$ are precisely $S_0 = Z_0 = 10$,
with $W_0(S_0,Z_0) = 10^{-11}$. In table 1 we report the exact VEV's and $F$--terms of the fields as numerically computed in the full model and their relative shifts compared to those computed, again numerically, in the simple effective model, with $S$ and $Z$ frozen at their values $S_0$ and $Z_0$.\footnote{For simplicity of the numerical analysis, we have not integrated out the massive $U(1)_X$ gauge field, as it should be done in both theories. However, this is not going to affect the vacuum or the mass spectrum and hence the results that follow are reliable.}
The model above belongs to the general class of models studied in the main text with an  $\epsilon$ roughly  ${\cal O}(10^{-13})$.
Keeping two significant digits, the physical masses are:
\bea
m_{H_1}^2 & = &  1.1\cdot 10^{-2}, \ \ \ \ \ m_{H_2}^2 = 1.0\cdot 10^{-2}, \ \ \ \ \   m_{\phi_r}^2  = 1.5\cdot 10^{-3}, \\
m_{T_i}^2 & = &  4.0\cdot 10^{-27}, \ \ \ \ 
m_{T_r}^2 =   3.7\cdot 10^{-27}, \ \ \ \ \ 
m_{\chi_i}^2 =2.3\cdot 10^{-28}, \ \
m_{\chi_r}^2 =2.3\cdot 10^{-28}. \nn
\eea
where $H_1\simeq Z+S$, $H_2\simeq Z-S$ and the subscript $r$ and $i$ denote real and imaginary field components, respectively. The masses $m_{H1}^2$ and $m_{H2}^2$ refer to both components of the complex scalar fields, being SUSY breaking effects negligible. 
The imaginary component of $\phi$ is approximately the Goldstone boson eaten by $A_X$ and hence is exactly massless. The relative mass shifts are
\bea
\Delta m_{\phi_r}^2 &  =  & -4.7 \cdot 10^{-15}\,, \ \ \ \ \ \ \
\Delta m_{T_i}^2 = -2.1 \cdot 10^{-14}\,, \ \
\Delta m_{T_r}^2= -2.6 \cdot 10^{-14}\,, \nn \\
\Delta m_{\chi_i}^2 & = &   1.4 \cdot 10^{-14}\,,\ \ \ \ \ \ \ \ \ 
\Delta m_{\chi_r}^2 =  1.8 \cdot 10^{-14}\,.
\eea
Finally, we also report the gravitino mass, the $D_X$--term, the cosmological constant and their relative shifts:
\bea
m_{3/2}^2 &  = &  9.5\cdot 10^{-31} \,, \ \ \  \Delta m_{3/2}^2 =-2.7 \cdot 10^{-14}\,, \nn \\
D_X & = & 3.7 \cdot 10^{-27}  \,, \, \  \ \ \ \Delta D_X = 1.5\cdot 10^{-14}\,, \nn \\
V_0 & = &  1.2 \cdot 10^{-32}\,,  \ \ \ \  \ \ \Delta V_0 = 8.2\cdot 10^{-12}\,.
\eea
Notice that being the cosmological constant fine-tuned to be ``small", namely of order  $10^{-2} m_{3/2}^2$,
its relative shift is larger. The latter is inversely proportional to the smallness of $V_0$.

\TABLE[t]{
\begin{tabular}{|c|c|c|c|c|c|c| }
\hline&
$\langle X \rangle $ & $\Delta \langle X \rangle$ &$F^X$ &$\Delta F^X$ \\ \hline
$S$ & $10+2\cdot 10^{-13}$ & $2\cdot 10^{-14}$ &  $-9.9\cdot 10^{-28}$ & --  \\
$Z$ & $10+2\cdot 10^{-13}$ &   $2\cdot 10^{-14}$ & $5.6\cdot 10^{-28}$ & -- \\
$T$ & $31.4$ &   $1.1\cdot 10^{-15}$  & $-3.4\cdot 10^{-15}$ &  $-1.4\cdot 10^{-16}$  \\
$\phi$ &  $0.15$ &  $-1.3\cdot 10^{-15}$  & $1.4\cdot 10^{-16}$ &  $ -7.4\cdot 10^{-14}$ \\
$\chi$ &  $0.05$ &  $-7.6\cdot 10^{-14}$  & $7.5\cdot 10^{-15}$ &  $1.5\cdot 10^{-14}$ \\
 \hline
\end{tabular}
\caption{VEV's and $F^N=e^{K/2}g^{\bar MN } \bar F_{\bar M}$ terms for the fields and their relative shifts, as derived by a numerical analysis.
Here and in the main text $\Delta X  \equiv (X_{full}-X_{sim})/X_{full}$. All quantities are in reduced Planck units.}
}

The relative shifts of the various quantities considered are smaller than $\epsilon$
due to the fact that the K\"ahler mixing between the $H$ and $L$ fields
coming from  (\ref{K4sugra}) in the above vacuum are relatively small.  This example shows the excellent agreement between the full and the simple theory.

\end{document}